\documentclass[preprints,article,accept,moreauthors,pdftex]{mdpi}
\firstpage{1} 

\pubvolume{0}
\issuenum{0}
\articlenumber{0}
\pubyear{2022}
\copyrightyear{2022}
\datereceived{} 
\dateaccepted{} 
\datepublished{} 
\hreflink{https://doi.org/} 
\newcommand{\Mpc}{\rm{~km~s^{-1}~Mpc^{-1}}}

\Title{Cosmological Parameter Estimation Using Current and Future Observations of Strong Gravitational Lensing}

\TitleCitation{Cosmological Parameter Estimation Using Current and Future Observations of Strong Gravitational Lensing}

\Author{Jing-Zhao Qi $^{1}$\orcidA{}, Wei-Hong Hu $^{1}$, Yu Cui $^{1}$, Jing-Fei Zhang $^{1}$\orcidC{} and Xin Zhang $^{1,2,3}$*\orcidB{}}

\AuthorNames{Jing-Zhao Qi, Wei-Hong Hu, Yu Cui, Jing-Fei Zhang and Xin Zhang}

\AuthorCitation{Qi, J.Z.; Hu, W.H.; Cui, Y.; Zhang, J.F.; Zhang, X.}

\address{%
$^{1}$ \quad  Department of Physics, College of Sciences, Northeastern University, Shenyang 110819, China\\
$^{2}$ \quad  Frontiers Science Center for Industrial Intelligence and Systems Optimization, Northeastern University, Shenyang 110819, China\\
$^{3}$ \quad  Key Laboratory of Data Analytics and Optimization for Smart Industry (Northeastern University), Ministry of Education, China\\}

\corres{Correspondence: zhangxin@mail.neu.edu.cn}

\abstract{
Remarkable development of cosmology is benefited from the increasingly improved measurements of cosmic distances including absolute distances and relative distances. In recent years, however, the emerged cosmological tensions motivate us to explore the independent and precise late-universe probes. The two observational effects of strong gravitational lensing (SGL), the velocity dispersions of lens galaxies and the time delays between multiple images, can provide measurements of relative and absolute distances respectively, and their combination is possible to break the degeneracies between cosmological parameters and enable tight constraints on cosmological parameters. In this paper, we combine the observed 130 SGL systems with velocity-dispersion measurements and 7 SGL systems with time-delay measurements to constrain dark-energy cosmological models. It is found that the combination of the two effects does not significantly break the degeneracies between cosmological parameters as expected. However, with the simulations of 8000 SGL systems with well-measured velocity dispersions and 55 SGL systems with well-measured time delays based on the forthcoming LSST survey, we find that the combination of two effects can significantly break the parameter degeneracies, and make the constraint precision of cosmological parameters meet the standard of precision cosmology. We conclude that the observations of SGL will become a useful late-universe probe for precisely measuring cosmological parameters.
}


\begin{document}
\section{Introduction}

The remarkable development of cosmology is benefited from the increasingly improved measurements of cosmic distances as a function of redshift. For instance, precise measurements of anisotropies in cosmic microwave background (CMB) radiation provide tight constraints on the acoustic horizon scale corresponding to the distance that sound waves have travelled till the last scattering, which enables the constraints on several fundamental cosmological parameters to be achieved with breathtaking precision \citep{WMAP:2003ivt,WMAP:2003elm,Aghanim:2018eyx}. Type Ia supernovae (SNe Ia) as standard candles could provide the relative luminosity distances, and the measurements of them led to the discovery of dark energy \citep{SupernovaSearchTeam:1998fmf,SupernovaCosmologyProject:1998vns}. In addition, the baryon acoustic oscillation (BAO) as the standard ruler can be used to determine angular diameter distances \citep{SDSS:2005xqv,BOSS:2016wmc}. However, it should be noted that the inherent scale for BAO standard ruler needs to be calibrated by CMB. Therefore, strictly speaking, the measurements of BAO are essentially also the relative distances.

The measurement of absolute distances is difficult, but it is important because the determination of one of the most fundamental parameters in cosmology, the Hubble constant $H_0$, is closely related to it. Moreover, recently, the measurement inconsistencies associated with $H_0$ are posing a serious challenge to modern cosmological theory \citep{DiValentino:2021izs,Vagnozzi:2019ezj,Zhang:2019ylr,Qi:2019zdk,Guo:2018uic,Vattis:2019efj,Guo:2019dui,Zhang:2014ifa,Guo:2018ans,Zhao:2017urm,Guo:2017qjt,Feng:2019jqa,Liu:2019awo,Zhang:2019cww,Ding:2019mmw}. The measurements of CMB power spectra by the \textit{Planck} satellite infer the value of $H_{0}=67.4\pm0.5~\rm{km\ s^{-1}\ Mpc^{-1}}$ assuming a flat $\Lambda$ cold dark matter ($\Lambda$CDM) model. However, in the local universe, the measured absolute distances of SNe Ia calibrated by distance ladders yield a larger value of $H_{0}=74.03\pm1.42$ $\rm km\ s^{-1}\ Mpc^{-1}$ \citep{Riess:2019cxk}. Above 4$\sigma$ tension between the values from two independent methods cannot be attributed to systematic errors crudely \citep{DiValentino:2018zjj,Riess:2019cxk}. In this context, the Hubble tension further highlights the importance of independent and precise measurements on absolute distances.

Strong gravitational lensing (SGL) can be used to measure an alternative absolute distance, the so-called time-delay distance $D_{\Delta t}$, which is a combination of three angular diameter distances between observer, lens, and source. Moreover, with the stellar velocity-dispersion measurements of the lens galaxy, the angular diameter distance $D_{\rm{l}}$ from observer to lens can also be obtained. In this way, the H0LiCOW ($H_{0}$ Lenses in COSMOGRAIL's Wellspring) team presented a measured value of $H_0=73.3^{+1.7}_{-1.8}~\rm{km\ s^{-1}\ Mpc^{-1}}$ with a 2.4\% precision from the observations of time delay for six lensed quasars \citep{Wong:2019kwg}. However, due to many difficulties in measuring the time delays of lensed quasars, there are only 7 observed samples at present \citep{Millon:2019slk,Rusu:2019xrq,Chen:2019ejq,DES:2019fny}. Even for the future surveys, like the Large Synoptic Survey Telescope (LSST) with wide field-of-view and frequent time sampling to monitor the SGL systems for time-delay measurements \citep{LSST:2008ijt,Huber:2019ljb}, are predicted to observe only a few dozen samples of well-measured SGL time delay \citep{Jee:2015yra,Wen:2019yem,Shajib:2017omw}. The small sample size makes it difficult to use as a powerful statistical quantity to precisely constrain cosmology.

In fact, for the galaxy-scale SGL samples, in addition to the time-delay measurements, several observed quantities can be used as the statistical quantities to constrain cosmological parameters, including the distribution of image angular separations \citep{Dev:2003fv,Cao:2012ja}, the distribution of lens redshifts \citep{Mitchell:2004gw,Cao:2011aq,Ma:2019cvt}, and the velocity dispersion of lens galaxies \citep{Biesiada:2006zf,Grillo:2007iv,Schwab:2009nz,Cao:2015qja}. Recently, using the lens velocity dispersion as statistical quantity in cosmology yields a series of achievements \citep{Qi:2018aio,Li:2018hyr,Cui:2017idf,Cao:2017nnq,Wang:2019yob,Wang:2022rvf,Wei:2022rcb,Geng:2021tiz,Chen:2018jcf,Liu:2021xvc,Qi:2022sxm}. With spectroscopic and astrometric data, 161 available samples are obtained with well-defined selection criteria at present \citep{Chen:2018jcf}. {Moreover, according to the predictions of the LSST survey \citep{Oguri:2010ns,Collett:2015roa,Goldstein:2018bue,Wojtak:2019hsc}, more than $1\times 10^5$ lensed galaxies and more than $8\times 10^3$ lensed quasars could potentially be observed.} Such a large sample is bound to yield extensive cosmological applications. For this method, the core idea is that the gravitational mass $M_{\rm{grl}}^{\rm{E}}$ equals to the dynamical mass $M_{\rm{dyn}}^{\rm{E}}$ within the Einstein radius $\theta_{\rm{E}}$ \citep{Chen:2018jcf}. The inferences of both masses are related to the cosmological distances, and the final formula derived from the equality of the two masses is a function of the distance ratio, $D_{\rm{ls}}/D_{\rm{s}}$, where $D_{\rm{ls}}$ is the angular diameter distance between lens and source, and $D_{\rm{s}}$ is the one between observer and source. In other words, the measurement provided by this method is relative distance.

In previous studies \citep{Cao:2011bg,Cao:2014jza,Cao:2015qja,Li:2015sla,Qi:2020rmm,Cao:2021zpf,Wu:2022dgy}, the respective applications of the two observed effects of SGL described above in cosmological parameter constraints have been fully discussed, as well as the improvements for parameter constraints by combining with CMB data \citep{Suyu:2013kha,Jee:2015yra,Shajib:2017omw}. Nevertheless, two incentives encourage us to improve and develop it further. Firstly, instead of the dependence on CMB as the precise early-universe probe to precisely constrain cosmology, to develop independent and precise late-universe probes is of great significance in the context of cosmological tensions indicating the inconsistencies between the early and late universe \citep{Verde:2019ivm}. Secondly, the determination of $H_0$ depends on the measurement of absolute distance, while the measurement of relative distance is helpful for constraint on other cosmological parameters such as the present matter density $\Omega_{m}$ and the equation of state of dark energy $w$ \citep{Cao:2011bg,Cao:2014jza,Cao:2015qja,Li:2015sla}. The combination of two independent observed effects of SGL providing the measurements of absolute distances and relative distances respectively is expected to break the cosmological parameter degeneracies and give tight constraints on them. In this paper, we will investigate what improvement the combination of these two observations will have on the cosmological constraint precision and whether a large sample of well-measured SGL data in the future LSST era can be used as a precise late-universe probe. Here, we consider three typical dark energy models for this analysis, i.e., the $\Lambda$CDM model, the $w$CDM model, and the Chevalliear-Polarski-Linder (CPL) model \citep{Linder:2002et}.
\section{Methodology and Data}
\label{sec:method}

\subsection{Velocity Dispersion of Lens Galaxies}
{The key point of using the lens velocity dispersion (VD)} as statistical quantity to constrain cosmological parameters is that the inferred gravitational mass $M_{\mathrm{grl}}^{\mathrm{E}}$ within the Einstein radius is equal to the projected dynamical mass $M_{\mathrm{dyn}}^{\mathrm{E}}$, namely $M_{\mathrm{grl}}^{\mathrm{E}}=M_{\mathrm{dyn}}^{\mathrm{E}}$. With the observations of angular separations between multiple images, the gravitational mass within the Einstein radius $\theta_{\rm{E}}$ could be inferred by
\begin{eqnarray}
\begin{array}{rr}
M_{\mathrm{grl}}^{\mathrm{E}}=\frac{c^{2}}{4 G} \frac{D_{\mathrm{s}} D_{\mathrm{l}}}{D_{\mathrm{ls}}} \theta_{\mathrm{E}}^{2},
\end{array}
\label{eq:Mgrl}
\end{eqnarray}
where $D_{\rm{s}}$ is the angular diameter distance between observer and source, $D_{\rm{l}}$ is that between observer and lens, and $D_{\rm{ls}}$ is that between lens and source.

On the other hand, assuming a mass distribution model for the lens galaxy, the projected dynamical mass $M_{\mathrm{dyn}}^{\mathrm{E}}$ can be derived. 
{The mass distribution of the lens galaxy is closely related to the constraints on cosmological parameters. The actual mass distribution of the lens galaxy is not necessarily an axisymmetric distribution, but a more ellipsoidal non-axisymmetric distribution. The isothermal elliptical model proposed by Kormann et al. \cite{kormann1994isothermal} is one of the common gravitational lensing models, which has caustics and critical curves in analytical form. Furthermore, ellipsoid models allow an estimation of effects due to the galaxy shape, and it fits well with mass profiles implied by observations \cite{Asada:2002je}. However, elliptical matter distributions are in general more difficult to handle, and here we take a simpler assumption of spherical symmetry. In this paper, we choose a general mass model for the lens galaxies \cite{Koopmans:2005ig,Chen:2018jcf}:}
\begin{align}
\label{eq:profile}
\begin{cases}
\rho(r)= \rho_0\; (r/r_0)^{-\gamma},\\
\nu(r)= \nu_0\; (r/r_0)^{-\delta},\\
\beta_{\textrm{ani}}(r)=1-\sigma_{\theta}^2/ \sigma_r^2,\\
\end{cases}
\end{align}
where $\rho(r)$ is the total mass density distribution, and $\nu(r)$ represents the luminous mass density distribution. The parameter $\beta_{\textrm{ani}}(r)$ characterizes the anisotropy of the stellar velocity dispersion, while $\sigma_{\theta}$ and $\sigma_{r}$ are the tangential and radial velocity dispersions, respectively. For the total mass density slope $\gamma$, according to the analysis in Ref. \citep{Chen:2018jcf}, the dependencies of $\gamma$ on both the redshift and the surface mass density should be taken into account. In this paper, therefore, we also adopt the parameterization of $\gamma$ as \citep{Chen:2018jcf}
\begin{equation}
\gamma=\gamma_0+\gamma_z z_{\rm{l}}+\gamma_s\log\tilde{\Sigma},
\end{equation}
where $\gamma_0$, $\gamma_z$ and $\gamma_s$ are constants, and $z_{\mathrm{l}}$ is the redshift of lens. Here, $\tilde{\Sigma}$ is the normalized surface mass density, defined as
\begin{eqnarray}
\tilde{\Sigma}=\frac{(\sigma_{0}/100 \textrm{km}\, \textrm{s}^{-1})^2}{R_{\textrm{eff}}/10h^{-1}\textrm{kpc}},
\end{eqnarray}
where $h=H_0/(100~\rm{km~s^{-1}~Mpc^{-1}})$, $R_{\rm{eff}}$ is the half-light radius of the lens galaxy, and $\sigma_0$ is the velocity dispersion of the lens galaxy.

For the luminous mass density slope $\delta$, it is commonly considered as a universal parameter for all lens galaxies in the entire sample. In fact, however, the individual value of $\delta$ for lens galaxies can be obtained by fitting the two-dimensional power-law luminosity profile over a circle of radius $\theta_{\rm{eff}}/2$ for lens galaxies with the high-resolution imaging data. In this way, a sample including 130 SGL systems with the $\delta$ observation is obtained \citep{Chen:2018jcf}. Moreover, \citet{Chen:2018jcf} concluded that the intrinsic scatter of $\delta$ among the lenses should be taken into account to get an unbiased cosmological estimate.

By combining the mass distribution model in Eq. (\ref{eq:profile}) and the well-known spherical Jeans equation, the total mass contained within a sphere with radius $r$ can be expressed as
\begin{equation}
M(r)=\frac{2}{\sqrt{\pi}}\frac{\Gamma(\gamma/2)}{\Gamma(\frac{\gamma-1}{2})}\left(\frac{r}{R_{\textrm{E}}}\right)^{3-\gamma}M^{\rm E}_{\textrm{dyn}},
\label{eq:Mr2}
\end{equation}
where the Einstein radius $R_{\mathrm{E}}$ is determined by $R_{\mathrm{E}}=D_{\mathrm{l}}\theta_{\mathrm{E}}$ \cite{Schwab:2009nz}, and $\Gamma(x)$ is Euler's Gamma function. The radial velocity dispersion $\sigma_{r}$ is determined by
\begin{equation}
\sigma_r^2(r)
=
\frac{2}{\sqrt{\pi}}\frac{GM_{\mathrm{dyn}}^{\mathrm{E}}}{R_{\mathrm{E}}}\frac{1}{\xi-2\beta_{\textrm{ani}}}\frac{\Gamma(\gamma/2)}
{\Gamma(\frac{\gamma-1}{2})}\left(\frac{r}{R_{\mathrm{E}}}\right)^{2-\gamma},
\label{eq:sigma_2r_2}
\end{equation}
where $\xi=\gamma+\delta-2$, and $\beta_{\textrm{ani}}$ is assumed to be independent of the radius $r$.

From the spectroscopic data, the velocity dispersion $\sigma_{\textrm{ap}}$ inside the circular aperture with the angular radius $\theta_{\textrm{ap}}$ could be measured. However, to consider the effect of the aperture size on the measurements of velocity dispersions, the velocity dispersion $\sigma_{\textrm{ap}}$ measured within certain apertures $\theta_{\textrm{ap}}$ should be normalized to a typical physical aperture via
\begin{eqnarray}
\label{eq:sigma_obs}
\sigma_{0}= \sigma_{\textrm{ap}}[\theta_{\textrm{eff}}/(2\theta_{\textrm{ap}})]^{\eta}.
\end{eqnarray}
According to Refs. \citep{Jorgensen:1995zz,Koopmans:2005ig,Chen:2018jcf}, the value of correction factor $\eta$ we adopt is $\eta = -0.066\pm0.035$. Based on the above analysis, the velocity dispersion could be expressed as
\begin{eqnarray}
\sigma_{0}=\sqrt{\frac{c^2}{2\sqrt{\pi}}\frac{D_{\rm{s}}}{D_{\rm{ls}}}\theta_{\rm{E}}F(\gamma,\delta,\beta_{\rm{ani}})\left(\frac{\theta_{\rm{eff}}}{2\theta_{\rm{E}}}\right)^{2-\gamma}},
\label{eq:sigma_th}
\end{eqnarray}
where
\begin{align}
F(\gamma,\delta, \beta_{\textrm{ani}})=&\frac{3-\delta}{(\xi-2\beta_{\textrm{ani}})(3-\xi)} \notag \\
&\times \left[\frac{\Gamma
\left[(\xi-1)/2\right]}{\Gamma(\xi/2)}-\beta_{\textrm{ani}}\frac{\Gamma\left[(\xi+1)/2\right]}{\Gamma\left[(\xi+2)/2
\right]}\right] \notag \\
& \times\frac{\Gamma(\gamma/2)\Gamma(\delta/2)}{\Gamma\left[(\gamma-1)/2\right]\Gamma\left[(\delta-1)/2\right]}.
\end{align}
For the detailed derivation and description, we refer the reader to Refs. \citep{Koopmans:2005ig,Chen:2018jcf}.

Considering the extra mass contribution from matters along the line of sight (LOS), we take about 3\% fractional uncertainty on velocity dispersion as the systematic error $\Delta\sigma_0^{\rm{sys}}$ \citep{Jiang:2007jx}. Together with the statistical error propagated from the measurement error $\Delta\sigma_0^{\rm{stat}}$ and the error caused by the aperture correction $\Delta\sigma_0^{\rm{AC}}$, the total uncertainty of $\sigma_0$ could be given by
\begin{eqnarray}
\label{eq:err_sigma_e2}
 (\Delta\sigma_{0}^{\textrm{tot}})^2 = (\Delta\sigma_{0}^{\textrm{stat}})^2+(\Delta\sigma_{0}^{\textrm{AC}})^2+
 (\Delta\sigma_{0}^{\textrm{sys}})^2.
\end{eqnarray}

The cosmological parameters could be constrained by maximizing the likelihood $\mathcal{L}_{\rm{VD}}\propto \exp\left(-\chi^2_{\rm{VD}}/2\right)$, and here $\chi^2_{\rm{VD}}$ is constructed as
\begin{eqnarray}
\label{eq:chi2}
\chi_{\textrm{VD}}^{2}=\sum^{N}_{i=1}\left(\frac{\sigma^{\textrm{th}}_{0 ,i}-\sigma^{\textrm{obs}}_{0 ,i}}
{\Delta\sigma^{\textrm{tot}}_{0 ,i}}\right)^2,
\end{eqnarray}
where $N$ is the number of the data points.

The observational sample of the velocity dispersion used in this paper originally includes 161 galaxy-scale SGL systems compiled by \citet{Chen:2018jcf}. However, as mentioned above, the intrinsic scatter of $\delta$ among the lens galaxies should be taken into account to get an unbiased cosmological estimate. Therefore, the SGL sample we adopt is the truncated sample including 130 SGL systems with the observations of $\delta$. {The relevant information necessary to perform statistical analyses for estimating cosmological parameters includes redshifts of lenses ($z_{\rm{l}}$) and sources ($z_{\rm{s}}$), Einstein angle ($\theta_{\rm{E}}$), effective radius ($\theta_{\rm{eff}}$), aperture angular radius ($\theta_{\rm{ap}}$), measured velocity dispersion ($\sigma_{\rm{ap}}$), and measured luminous mass density slope ($\delta$).} The more detailed analyses and descriptions, we refer the reader to Ref. \citep{Chen:2018jcf}. For convenience, we use the abbreviation ``VD" to represent this SGL sample.

\subsection{Time-Delay Measurements}
For an SGL system, the emitted light rays from the background object (the source) corresponding to the different image positions pass different paths and gravitational potentials, which makes the time delays between the arrival times of the light rays. If the source is variable, time delays between multiple images can be measured by long-term dedicated photometric monitoring \citep{Schechter:1996fa,Fassnacht:1999re,Fassnacht:2002df,Kochanek:2005ge,Courbin:2010au}. The time delay between two images is related to both the time-delay distance and the gravitational potential of the lens galaxy via the relation:
\begin{align}
\Delta t_{ij}=\displaystyle{\frac{D_{\Delta t}}{c}}\left[\displaystyle{\frac{({\boldsymbol{\theta}}_i-\boldsymbol{\beta})^2}{2}}-\psi({\boldsymbol{\theta}}_i)-\displaystyle{\frac{({\boldsymbol{\theta}}_j-\boldsymbol{\beta})^2}{2}}+\psi({\boldsymbol{\theta}}_j)\right],
\end{align}
where $\boldsymbol{\theta}_{i}$ and $\boldsymbol{\theta}_{j}$ are the coordinates of the images $i$ and $j$ in the lens plane, respectively. The source position, $\boldsymbol{\beta}$, and lens potentials, $\psi\left(\boldsymbol{\theta}_{i}\right)$ and $\psi\left(\boldsymbol{\theta}_{j}\right)$, can be determined from the mass model of the system. With the measurements of the time delay $\Delta t$, the time-delay distance $D_{\Delta t}$ can be inferred, which is the combination of three angular diameter distances \cite{Refsdal:1964nw,Suyu:2009by,Bonvin:2016crt}:
\begin{eqnarray}
D_{\Delta t} \equiv\left(1+z_{\mathrm{l}}\right) \frac{D_{\mathrm{l}} D_{\mathrm{s}}}{D_{\mathrm{ls}}}.
\end{eqnarray}
It is important to note that the angular diameter distance $D_{\rm{l}}$ also could be obtained by combining the time-delay measurements with the stellar velocity dispersion measurements of the lens galaxy, which not only could improve the constraints on the cosmological parameters, but also is helpful to break the mass-sheet degeneracy \citep{Shajib:2017omw}. 

In Table~\ref{TD_data}, we summarize the existing seven SGL systems with measured time delays $D_{\Delta t}$ and the angular diameter distances $D_{\rm{l}}$. {The relevant information necessary to perform statistical estimation of cosmological parameters includes the redshifts of lens and source, the posterior distribution of $D_{\Delta t}$ and $D_{\rm{l}}$ in the form of Monte Carlo Markov chains (MCMCs). Here, it should be noted that a kernel density estimator is used to compute the posterior distribution $\mathcal{L}_{D_{\Delta t}}$ from MCMCs. The sampling software could be found in the website ({\url{https://doi.org/10.5281/zenodo.3633035}}) and the posterior distributions of $D_{\Delta t}$ and $D_{\rm{l}}$ in the form of MCMCs are available in the H0LiCOW website ({\url{http://www.h0licow.org}}). For convenience, we use the abbreviation ``TD'' to represent the time-delay measurements.}

For the angular diameter distances involved in VD and TD measurements, in the framework of a flat universe, their theoretical expresses $D_{\rm{l}}$, $D_{\rm{s}}$ and $D_{\rm{ls}}$ are given by
\begin{eqnarray}
D_{\rm{l}}(z_{\rm{l}};\textbf{p})=\frac{c}{H_0(1+z_{\rm{l}})}\int_0^{z_{\rm{l}}}\frac{dz}{E(z;\textbf{p})},\\
D_{\rm{s}}(z_{\rm{s}};\textbf{p})=\frac{c}{H_0(1+z_{\rm{s}})}\int_0^{z_{\rm{s}}}\frac{dz}{E(z;\textbf{p})},\\
D_{\rm{ls}}(z_{\rm{l}},z_{\rm{s}};\textbf{p})=\frac{c}{H_0(1+z_{\rm{s}})}\int_{z_{\rm{l}}}^{z_{\rm{s}}}\frac{dz}{E(z;\textbf{p})},
\end{eqnarray}
respectively. Here $E(z)\equiv H(z)/H_0$ is the dimensionless Hubble parameter, and $\textbf{p}$ denotes the parameters of the considered cosmological model.

\begin{table}[H] 
\caption{$D_{\Delta t}$ and $D_{\rm{l}}$ for the seven lenses.\label{TD_data}}
\newcolumntype{C}{>{\centering\arraybackslash}X}
\begin{tabularx}{\textwidth}{CCCCCC}
\toprule
\textbf{Lens name}	& \textbf{$z_{\rm{l}}$}	& \textbf{$z_{\rm{s}}$} & \textbf{$D_{\Delta t}$ (Mpc)} & \textbf{$D_{\rm{l}}$ (Mpc)} & \textbf{References}\\
\midrule
B1608+656      &   0.6304      &   1.394   &   $5156^{+296}_{-236}$& $1228^{+177}_{-151}$  & \cite{Suyu:2009by,Jee:2019hah}    \\
RXJ1131-1231   &   0.295       &   0.654   &   $2096^{+98}_{-83}$& $804^{+141}_{-112}$    & \cite{Suyu:2013kha,Chen:2019ejq}   \\
HE 0435-1223   &   0.4546      &   1.693   &   $2707^{+183}_{-168}$ & ---  & \cite{Wong:2016dpo,Chen:2019ejq}    \\
SDSS 1206+4332 &   0.745       &   1.789   &   $5769^{+589}_{-471}$& $1805^{+555}_{-398}$  & \cite{Birrer:2018vtm}    \\
WFI2033-4723   &   0.6575      &   1.662   &   $4784^{+399}_{-248}$& ---  & \cite{Rusu:2019xrq}   \\
PG 1115+080    &   0.311       &   1.722   &   $1470^{+130}_{-127}$& $697^{+186}_{-144}$  & \cite{Chen:2019ejq}   \\
DES J0408-5354 &   0.597       &   2.375   &   $3382^{+146}_{-115}$ & $1711^{+376}_{-280}$  & \cite{DES:2019fny,Agnello:2017mwu}    \\
\bottomrule
\end{tabularx}
\end{table}
\unskip


\section{Results and Discussions}

The observations of velocity dispersion for SGL systems provide the measurements of relative distances, while the time-delay observations could offer the absolute distances. In this section, we will present the constraints on cosmological models from these two observations derived from SGL systems to see whether they can break the degeneracy between cosmological parameters. Here, we use the \texttt{emcee} \citep{Foreman-Mackey:2012any} Python module based on the MCMC analysis to implement the cosmological constraints.

\subsection{The Constraints on Cosmological Parameters with Current Observations of SGL}

\begin{figure}[H]
\includegraphics[width=10.5 cm]{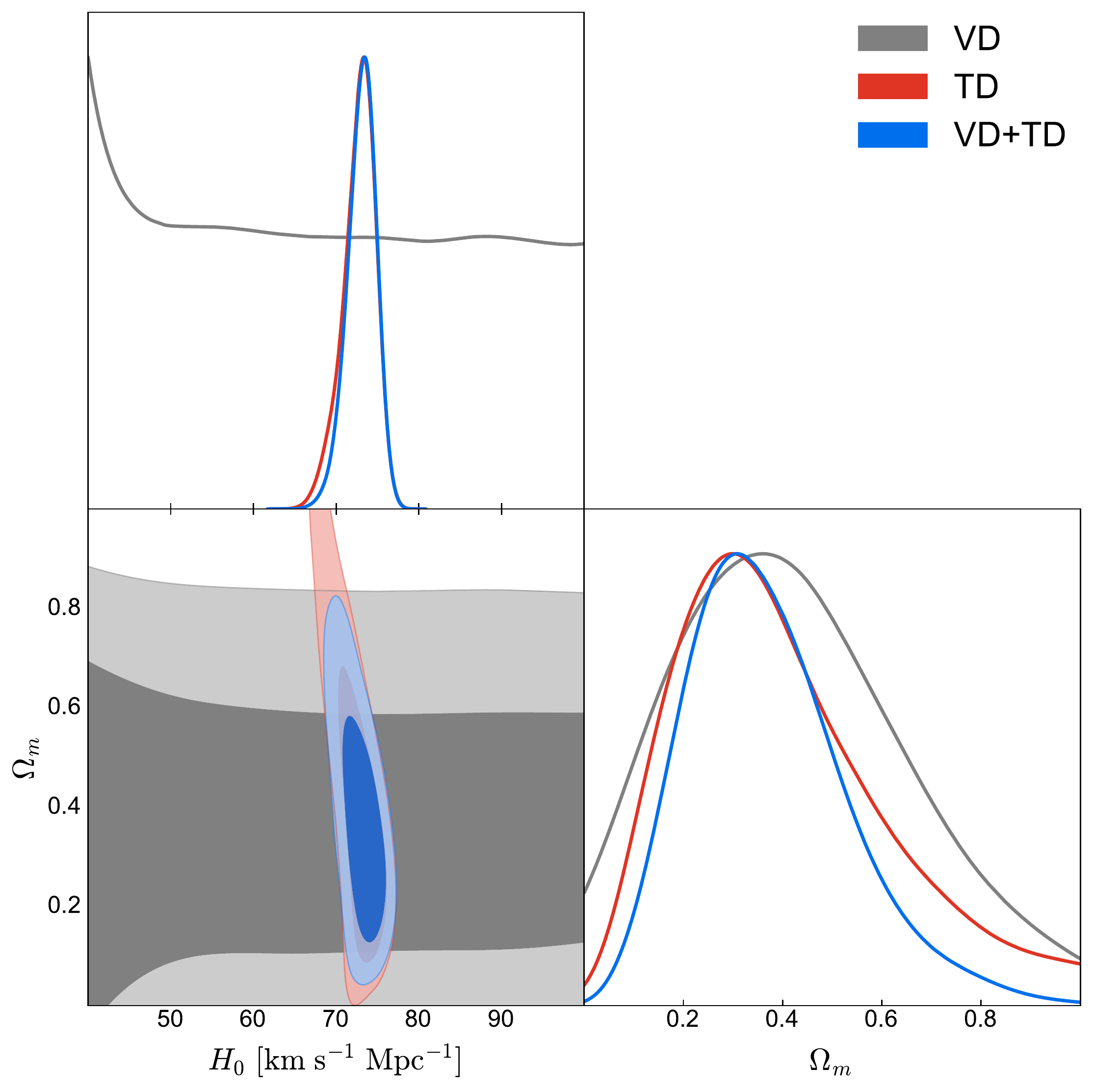}
\caption{The constraints (68.3\% and 95.4\% confidence level) on the $\Lambda$CDM model from VD, TD, and VD+TD. \label{lcdm}}
\end{figure}

In Figure~\ref{lcdm} and Table \ref{constraints}, we show the constraints on the $\Lambda$CDM model from VD, TD and the combination of them, i.e. VD+TD. We can see clearly that the constraint on $H_0$ from the VD observation is invalid due to the relative distance measurements. The constraint on $\Omega_m$ from VD is rather weak, $\Omega_m=0.400^{+0.256}_{-0.216}$. For the results from the TD observation, the constraint on $\Omega_m$ we get is $\Omega_m=0.362^{+0.247}_{-0.170}$, which is comparable with that from VD. Since TD could measure absolute distances, a tight constraint on $H_0$ is obtained, i.e., $H_0=72.99^{+1.72}_{-2.20}\Mpc$. On the other hand, we can see that the combination of VD and TD does not break significantly the degeneracy between cosmological parameters as expected. The constraint results from VD+TD are $\Omega_m=0.350^{+0.175}_{-0.139}$ and $H_0=73.20^{+1.60}_{-1.86} \Mpc$, from which we can see that the constraints are only slightly improved by the combination.

\begin{figure}[H]
\includegraphics[width=10.5 cm]{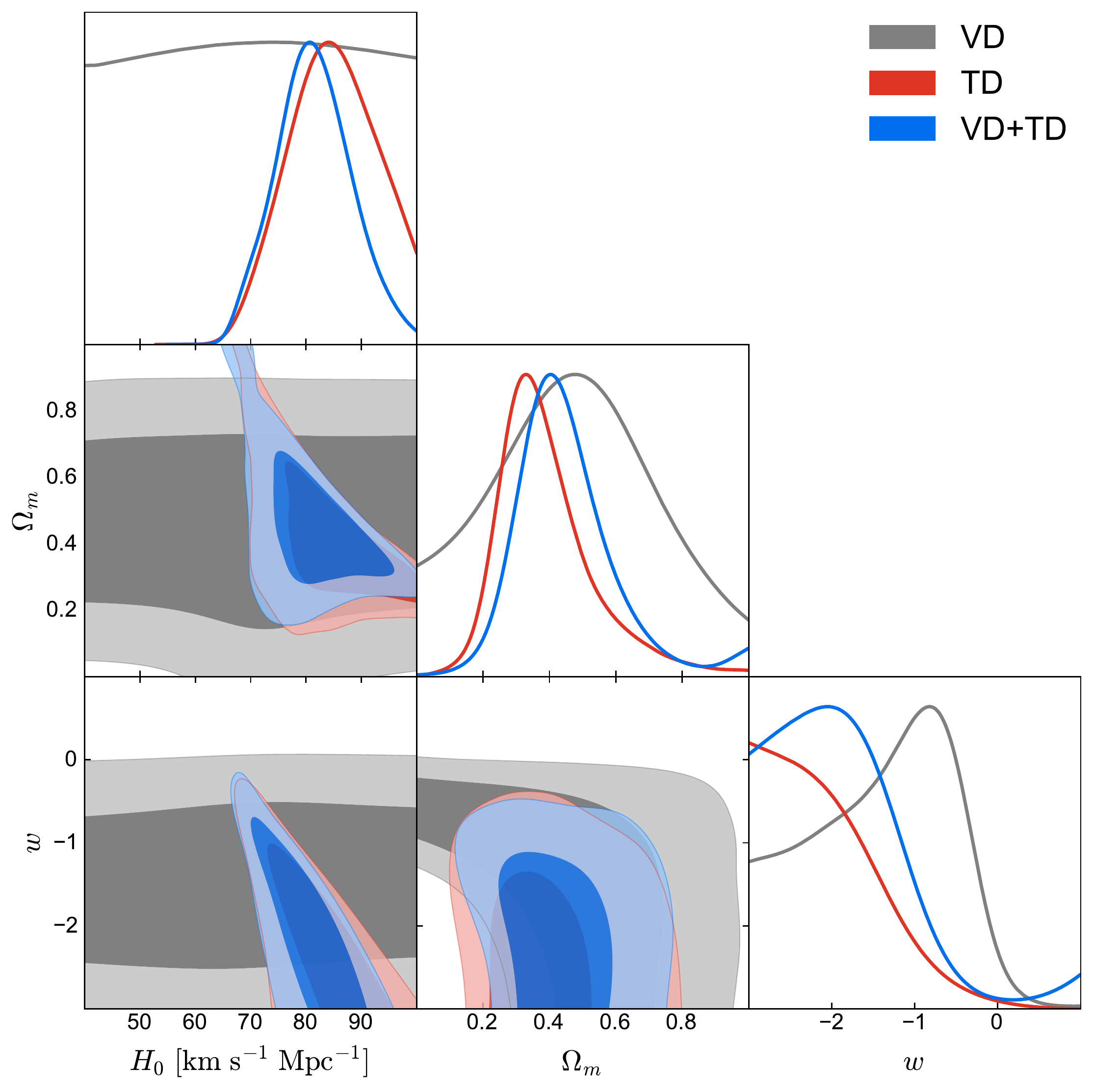}
\caption{The constraints (68.3\% and 95.4\% confidence level) on the $w$CDM model from VD, TD, and VD+TD.\label{wcdm}}
\end{figure}

We present the constraints on the $w$CDM model in Figure~\ref{wcdm} and Table \ref{constraints}. We find that the constraints on $\Omega_m$ from VD, TD and VD+TD are similar to those in the case of $\Lambda$CDM, for both best-fit values and constraint errors. For $H_0$, the TD data give $H_0=85.29^{+9.75}_{-8.37} \Mpc$. The combination VD+TD gives $H_0=81.27^{+7.23}_{-6.64}\Mpc$, and we can see that the data combination only provide a slight improvement because VD cannot effectively constrain $H_0$. From the posterior probability distribution of $w$ in Figure~\ref{wcdm}, we can see that the VD data could offer a tighter constraint than TD for the parameter $w$. The combination VD+TD gives $w=-2.33^{+0.96}_{-1.05}$. {We find that a phantom-type dark energy (with $w<-1$) is preferred by the VD+TD data, which leads to a high value of the Hubble constant.
Here we also notice that the VD+TD constraint on $w$ is rather weak and its best-fit value significantly deviates from the result $w=-1.03\pm0.03$ derived from CMB+BAO+SNe \citep{Aghanim:2018eyx}. This is mainly because the current sample sizes of VD and TD are rather small (7 TD data and 130 VD data). Such small-size SGL samples cannot accurately and precisely constrain the EoS of dark energy, and so the current result is much worse than that of CMB+BAO+SNe. On the other hand, the constraints on cosmological parameters depend on accurate modeling of lens models. The mass model of the lens galaxies we adopt is assumed to have a spherical symmetry. The deviation from reality may lead to a bias in estimates of cosmological parameters. Therefore, constructing a more reasonable lens model using future large samples and more accurate observational data is very important for cosmological parameter estimation.}

\begin{figure}[H]
\includegraphics[width=10.5 cm]{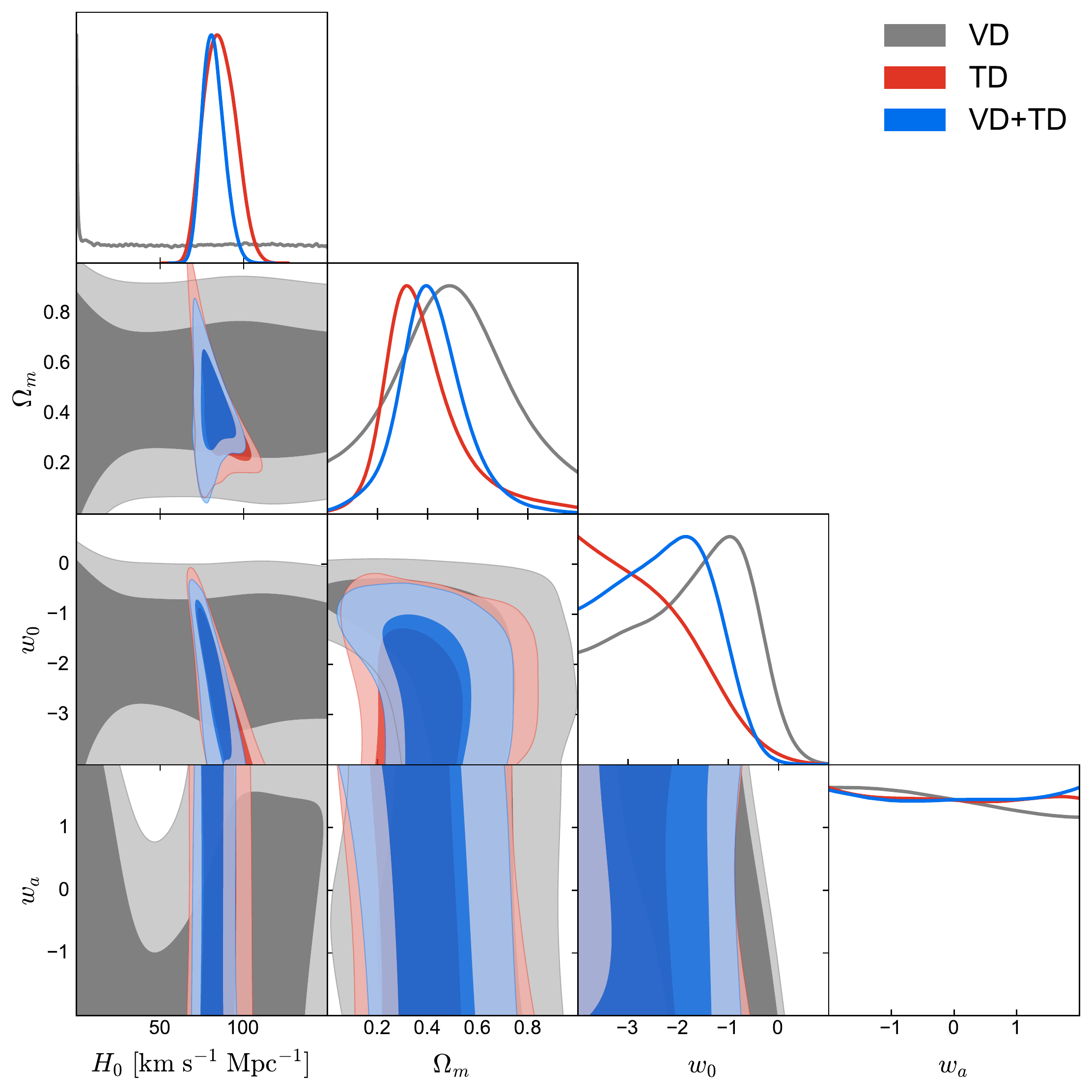}
\caption{The constraints (68.3\% and 95.4\% confidence level) on the CPL model from VD, TD, and VD+TD.\label{cpl}}
\end{figure}

For the CPL model, the constraint results from VD, TD, and VD+TD are shown in Figure~\ref{cpl} and Table \ref{constraints}. Compared with the case of $w$CDM, the best-fit values and constraint errors of $\Omega_m$, $H_0$, and $w_0$ are not changed a lot, even though one more free parameter is added. For $w_a$, we find that both VD and TD cannot give an effective constraint.

\unskip

\unskip

\unskip

\subsection{Forecast for the Constraints on Cosmological Parameters with the Future Observations}

In this subsection, we make a forecast for the constraints on cosmological parameters with the future SGL observations.

According to some estimates, the LSST will observe more than 8000 lensed quasars, about 3000 of which have well-measured time delays during the 10-year survey duration \citep{Oguri:2010ns,Collett:2015roa,Goldstein:2018bue,Wojtak:2019hsc}. Such a large sample is bound to bring a significant improvement for the estimation of cosmological parameters. Therefore, we perform a simulation for a realistic population of SGL. For the estimate of velocity dispersion, we simulate 8000 well-measured SGL systems with the 5\% and 1\% uncertainties for the observed velocity dispersion and Einstein radius, respectively, according to the analysis from Ref. \citep{Cao:2017nnq}. On the other hand, for the measurement of time delay, the sample requires accurate characterization for the mass distribution of the lens galaxy, auxiliary data such as high-resolution imaging, and stellar velocity dispersion observations. By selecting with strict criteria \citep{Jee:2015yra,Wen:2019yem}, there will be about 55 SGL systems with well-measured time-delay distance $D_{\Delta t}$ and angular diameter distances $D_{\rm{l}}$. According to the constraints on current lensed quasars \citep{Rusu:2019xrq,Chen:2019ejq,DES:2019fny,Suyu:2016qxx,Suyu:2013kha}, we set 5\% uncertainties for the time-delay measurements, 3\% for the lens mass modelling uncertainties, and 3\% for the lens environment uncertainties, all of which assign 6.6\% uncertainty to the time-delay distances for each SGL system \citep{Suyu:2020opl}. For the precision on $D_{\rm{l}}$,  we set 5\% uncertainty for it as in Refs. \citep{Jee:2015yra,Wen:2019yem,Suyu:2020opl}. In this simulation, we adopt the $\Lambda$CDM model as a fiducial model with the values of cosmological parameters $\Omega_m=0.315$ and $H_0=67.4\Mpc$ taken from Planck 2018 results \citep{Aghanim:2018eyx}. 

\begin{figure}[H]
\includegraphics[width=10.5 cm]{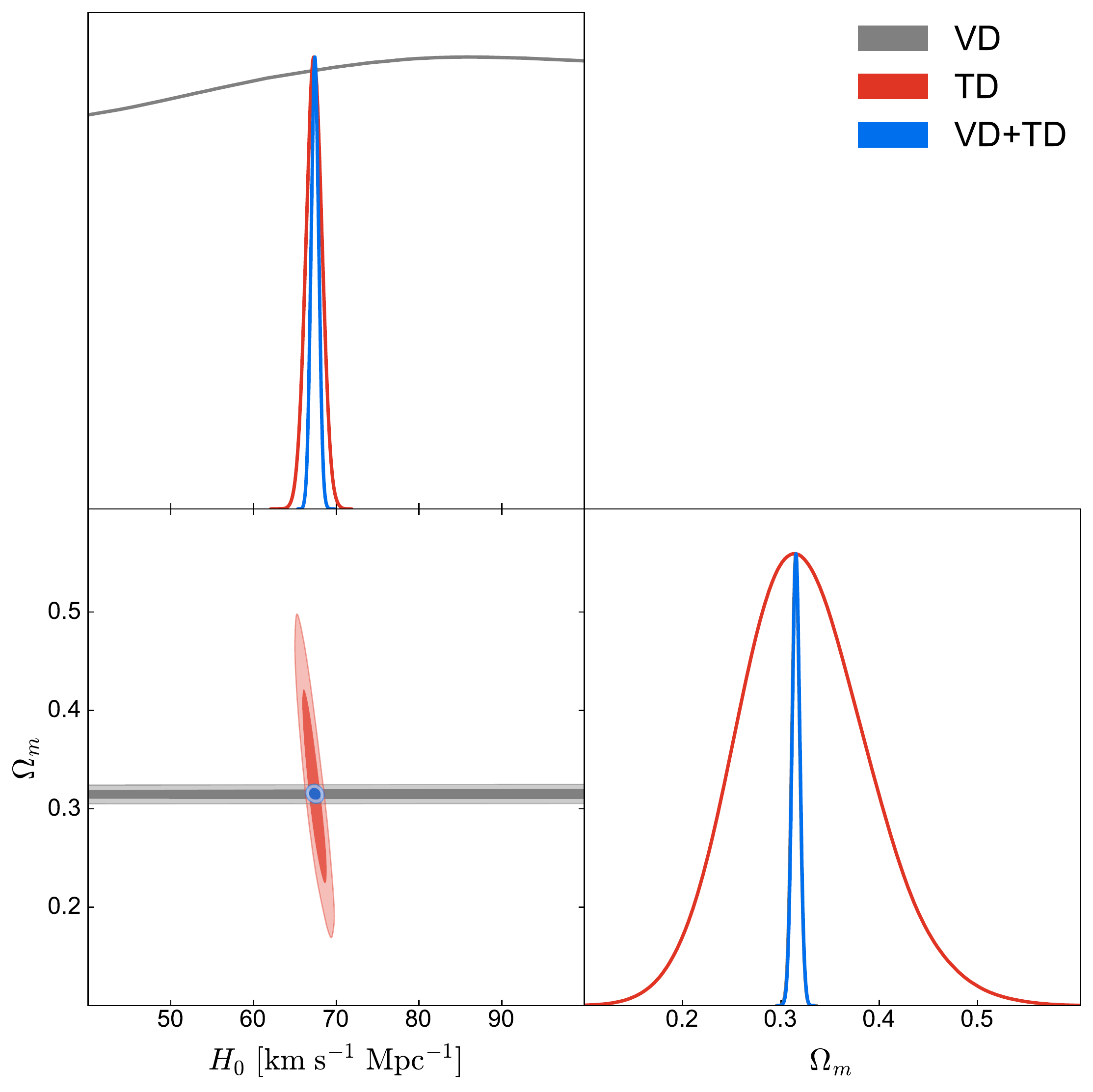}
\caption{The constraints (68.3\% and 95.4\% confidence level) on the $\Lambda$CDM model from the simulations of VD, TD, and VD+TD. \label{mclcdm}}
\end{figure}   
\unskip

\begin{figure}[H]
\includegraphics[width=10.5 cm]{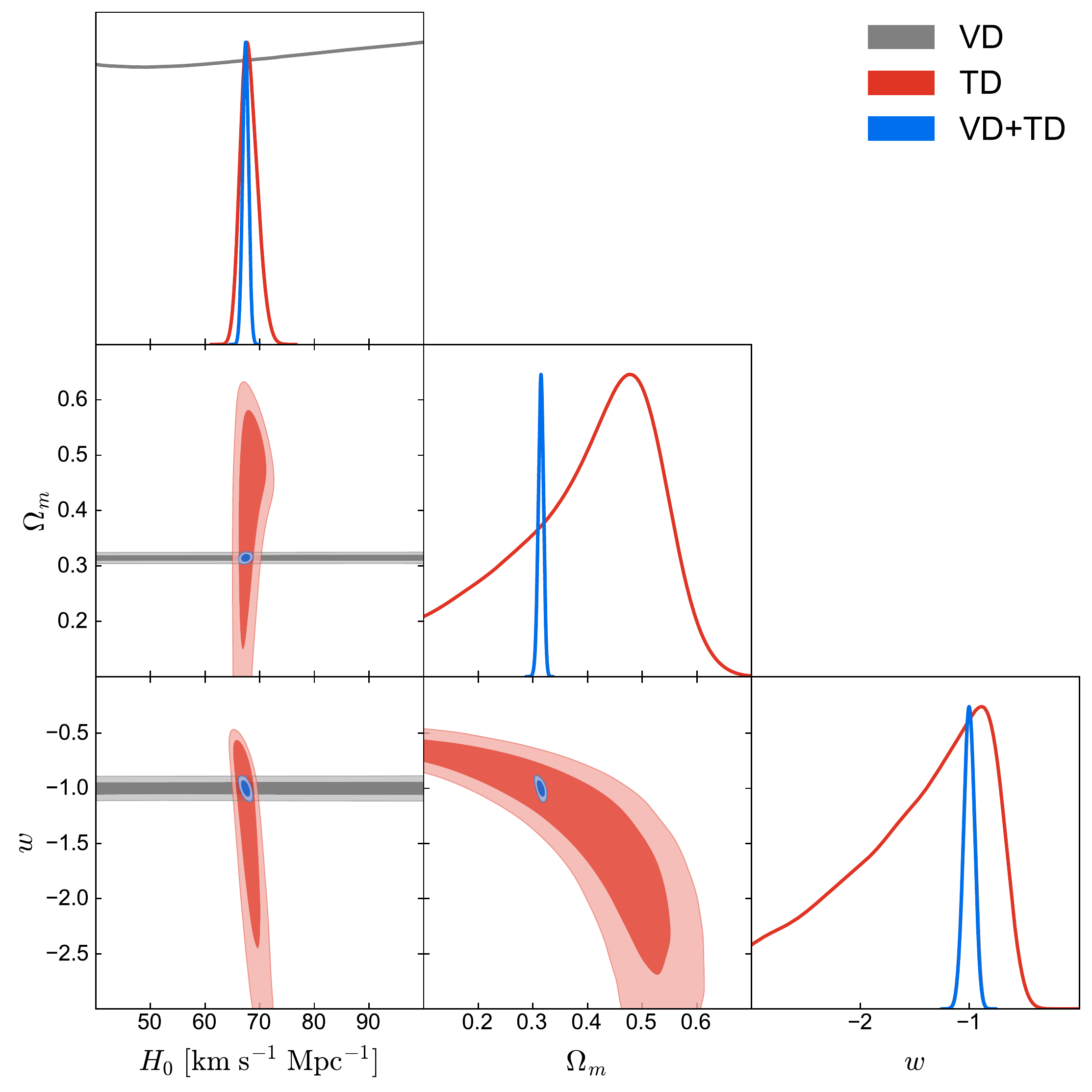}
\caption{The constraints (68.3\% and 95.4\% confidence level) on the $w$CDM model from the simulations of VD, TD, and VD+TD.\label{mcwcdm}}
\end{figure}   
\unskip

\begin{figure}[H]
\includegraphics[width=10.5 cm]{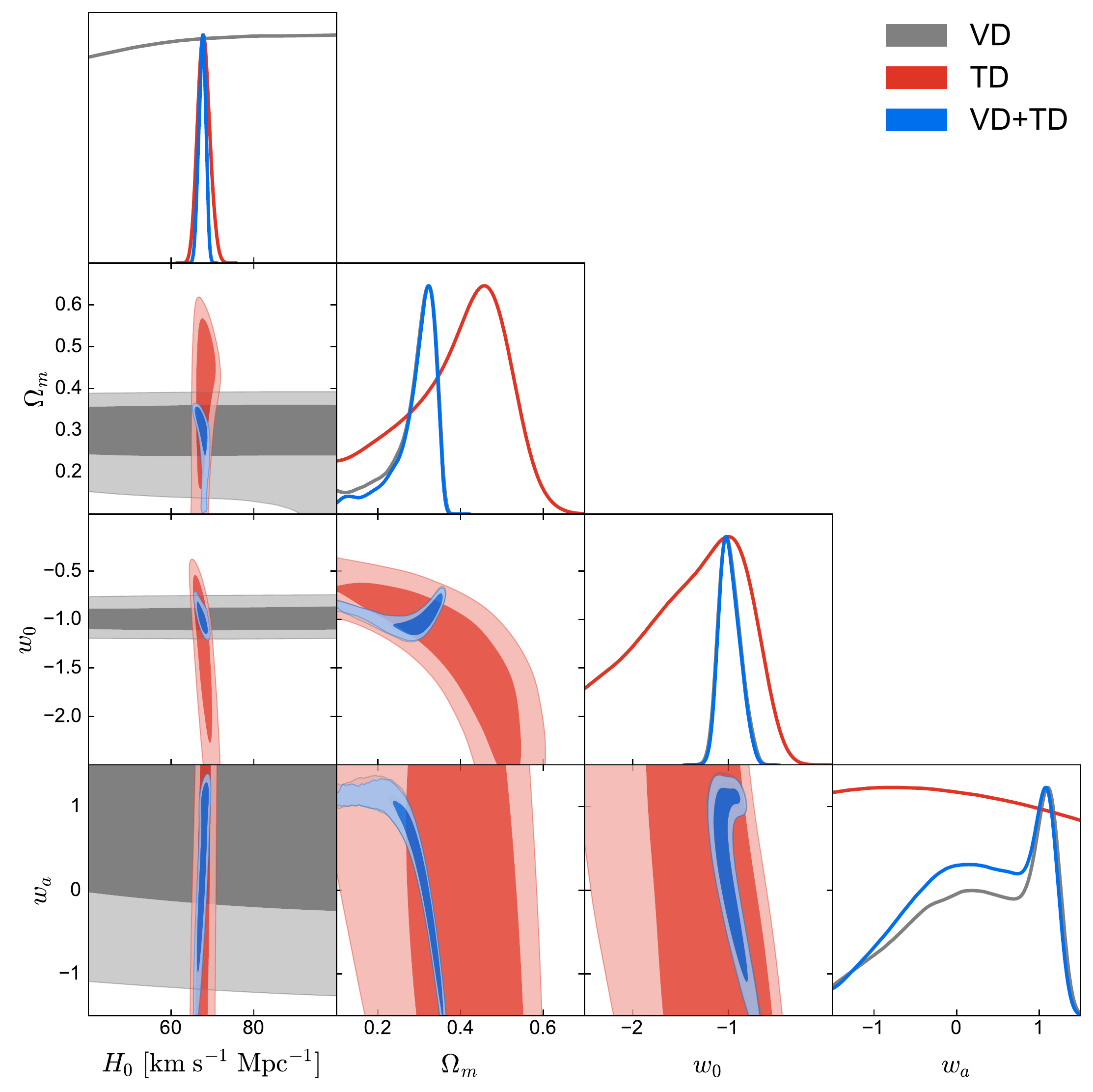}
\caption{The constraints (68.3\% and 95.4\% confidence level) on the CPL model from the simulations of VD, TD, and VD+TD.\label{mccpl}}
\end{figure}   
\unskip

\begin{table}[H] 
\caption{The constraints on cosmological parameters of the $\Lambda$CDM, $w$CDM and CPL models from the current VD, TD, and VD+TD. Here $H_{0}$ is in units of $\rm km\ s^{-1}\ Mpc^{-1}$.\label{constraints}}
\renewcommand{\arraystretch}{1.8}
\newcolumntype{C}{>{\centering\arraybackslash}X}
\begin{tabularx}{\textwidth}{CCCCC}
\toprule
\textbf{Model}	& \textbf{Parameter}	& \textbf{VD} & \textbf{TD}               & \textbf{VD+TD}  \\
\midrule
\multirow{2}{*}{$\Lambda$CDM}
    &$H_{0}$                    &$-$                          &$72.99^{+1.72}_{-2.20}$         &$73.20^{+1.60}_{-1.86}$  \\
    &$\Omega_{\rm m}$           &$0.400^{+0.256}_{-0.216}$    &$0.362^{+0.247}_{-0.170}$      &$0.350^{+0.175}_{-0.139}$
    \\ \midrule
\multirow{3}{*}{$w$CDM}
    &$H_{0}$                    &$-$                          &$85.29^{+9.75}_{-8.37}$         &$81.27^{+7.23}_{-6.64}$  \\
    &$\Omega_{\rm m}$           &$0.471^{+0.250}_{-0.253}$    &$0.365^{+0.159}_{-0.102}$      &$0.428^{+0.149}_{-0.112}$ \\
    &$w$                        &$-1.28^{+0.75}_{-1.07}$   &$-2.72^{+1.05}_{-0.89}$      &$-2.33^{+0.96}_{-1.05}$
    \\ \midrule
\multirow{4}{*}{CPL}
    &$H_{0}$                    &$-$                          &$85.30_{-9.26}^{+9.95}$     &$81.46^{+7.00}_{-6.17}$ \\
    &$\Omega_{\rm m}$           &$0.490_{-0.235}^{+0.232}$    &$0.359^{+0.175}_{-0.110}$      &$0.410^{+0.137}_{-0.116}$ \\
    &$w_0$                      &$-1.67^{+1.04}_{-1.46}$   &$-2.68^{+1.16}_{-0.93}$     &$-2.28_{-1.11}^{+0.96}$ \\
    &$w_a$                      &$-0.091^{+1.403}_{-1.296}$   &$-0.025^{+1.387}_{-1.359}$     &$0.010^{+1.367}_{-1.379}$ \\
\bottomrule
\end{tabularx}
\end{table}
\unskip

\begin{table}[H] 
\caption{The constraints on cosmological parameters of the $\Lambda$CDM, $w$CDM and CPL models by using the simulated VD, TD, and VD+TD. For comparison, we also list the fit results from the current CMB+BAO+SNe data, taken from Ref.~\citep{Zhang:2019loq}. Here $H_{0}$ is in units of $\Mpc$. \label{simulation}}
\renewcommand{\arraystretch}{1.5}
\newcolumntype{C}{>{\centering\arraybackslash}X}
\begin{tabularx}{\textwidth}{CCCCCC}
\toprule
\textbf{Model}	& \textbf{Parameter}	& \textbf{VD} & \textbf{TD} & \textbf{VD+TD} & \textbf{CMB+BAO+SNe} \\
\midrule
\multirow{4}{*}{$\Lambda$CDM}
            &$H_0$              &$-$           &$69.87_{-1.02}^{+1.00}$      &$69.99\pm 0.47$      &$67.64\pm 0.44$  \\
            &$\sigma(H_0)$      &$-$           &$1.01$      &$0.47$      &$0.44$ \\
            &$\Omega_{\rm m}$   &$0.300\pm 0.004$       &$0.310_{-0.062}^{+0.068}$     &$0.300\pm 0.004$     &$0.314\pm 0.006$ \\
            &$\sigma(\Omega_{\rm m})$   &$0.004$       &$0.065$     &$0.004$     &$0.006$  \\     
            \midrule
 \multirow{6}{*}{$w$CDM}
            &$H_0$              &$-$           &$70.67_{-1.52}^{+1.78}$       &$70.00_{-0.56}^{+0.57}$      &$67.90\pm 0.83$ \\
            &$\sigma(H_0)$              &$-$           &$1.65$       &$0.57$      &$0.83$ \\
            &$\Omega_{\rm m}$   &$0.300\pm 0.005$       &$0.418_{-0.160}^{+0.094}$     &$0.300_{-0.005}^{+0.004}$     &$0.312\pm 0.008$ \\
            &$\sigma(\Omega_{\rm m})$   &$0.005$       &$0.127$     &$0.005$     &$0.008$  \\
            &$w$                &$-1.00\pm 0.05$       &$-1.44_{-0.85}^{+0.56}$ &$-1.00\pm 0.05$     &$-1.01\pm 0.03$  \\           
            &$\sigma(w)$                &$0.05$       &$0.71$     &$0.05$     &$0.03$  \\
            \midrule
 \multirow{8}{*}{CPL}
            &$H_0$              &$-$        &$70.50_{-1.48}^{+1.61}$   &$70.22_{-0.96}^{+0.82}$   &$67.91\pm 0.83$ \\
            &$\sigma(H_0)$      &$-$        &$1.55$                     &$0.89$                    &$0.83$ \\
            &$\Omega_{\rm m}$   &$0.289_{-0.072}^{+0.032}$   &$0.403_{-0.155}^{+0.095}$   &$0.288_{-0.077}^{+0.031}$  &$0.312\pm 0.008$ \\
            &$\sigma(\Omega_{\rm m})$   &$0.052$        &$0.125$    &$0.054$     &$0.008$ \\
            &$w_0$              &$-1.00_{-0.10}^{+0.13}$  &$-1.36_{-0.66}^{+0.49}$   &$-1.00_{-0.09}^{+0.11}$   &$-0.99\pm 0.08$\\ 
            &$\sigma(w_0)$      &$0.12$        &$0.58$    &$0.10$     &$0.08$  \\ 
            &$w_a$   &$0.265_{-0.910}^{+0.795}$  &$-0.069_{-0.988}^{+1.053}$   &$0.288_{-0.865}^{+0.769}$  &$-0.10_{-0.27}^{+0.36}$ \\
            &$\sigma(w_a)$      &$0.853$         &$-$        &$0.817$     &$0.315$  \\
\bottomrule
\end{tabularx}
\end{table}
\unskip

The one-dimensional marginalized posterior distributions and the two-dimensional contours of parameters from VD, TD, and VD+TD are shown in Figures~\ref{mclcdm}--\ref{mccpl}. {The constraint results of the parameters are listed in Table \ref{simulation}. It can be seen that all best-fit values of parameters are consistent with the fiducial values at 1 $\sigma$ confidence level as expected. The key issue we wish to investigate is the constraint capability of the TD and VD observations in the future, which could be directly indicated by the constraint errors (1$\sigma$) of parameters.} We can see clearly that the combination VD+TD could effectively break the degeneracy between $H_0$ and $\Omega_m$ in these three models. In the $\Lambda$CDM model, the joint constraint gives the results $\sigma(H_0)=0.47 \Mpc$ and $\sigma(\Omega_m)=0.004$, and we find that the precisions have exceeded the Planck 2018 results, meeting the standard of precision cosmology. In the $w$CDM model, the constraint on $w$ from VD+TD is $\sigma(w)= 0.05$, which is comparable with the result of Planck 2018 TT,TE,EE+lowE+lensing+SNe+BAO, $\sigma(w)= 0.034$ \citep{Zhang:2019loq}. In fact, from the posterior distribution of $w$ in Figure~\ref{mcwcdm}, we find that the TD observation only give a very weak constraint on $w$, and the tight constraint is mainly contributed by the VD observation. In the CPL model, the TD observation does not provide an effective constraint on $w_a$, and even the combination VD+TD cannot offer an effective constraint on it. $\Omega_m$ and $w_0$ can be tightly constrained by the VD observation, which is benefited from the large sample size. On the other hand, although the TD observation can measure absolute distances, its ability to constrain cosmological parameters is very limited (except for $H_0$) due to its small sample size.

In summary, the TD observation can only constrain $H_0$ strictly but has weak ability to constrain other cosmological parameters. The VD observation is highly complementary with the TD observation, which could offer tight constraints on $\Omega_m$ and $w_0$. Therefore, we can see that the combination of VD and TD could significantly break the parameter degeneracy so that the constraint precisions meeting the standard of precision cosmology can be obtained. In the context of cosmological tensions between early and late universe, the observations of SGL provide an alternative way to precisely measure cosmological parameters in the late universe and explore the nature of dark energy.

\section{Conclusion}

The increasingly improved measurements of cosmic distances including relative distances and absolute distances enable the remarkable development of cosmology and the precise constraints on some fundamental cosmological parameters. Meanwhile, however, the measured discrepancies among several key cosmological parameters have emerged from the observations of early universe and late universe, which motivates us to explore the independent and precise late-universe probes. The two observed effects of SGL, time-delay measurements and the lens velocity dispersions, provide the measurements of absolute distances and relative distances, respectively, which are expected to break the cosmological parameter degeneracies and give tight constraints on them. In this paper, we combine the current observed 130 SGL systems with velocity dispersion and 7 SGL systems with time-delay measurements to constrain the $\Lambda$CDM, $w$CDM, and CPL models. We find that the TD observation is only sensitive to $H_0$ due to its absolute distance measurements. Moreover, the inference of $H_0$ from TD is strongly cosmological model dependent. In the $\Lambda$CDM model, the constraint on $H_0$ from the combination of VD and TD is $H_0=73.20^{+1.60}_{-1.86}\Mpc$, which is consistent with the results measured by local distance ladder from the SH0ES collaboration \citep{Riess:2020fzl}. Compared with the constraints from individual VD and TD, the combination of them does not break significantly the degeneracy between cosmological parameters as expected.

It is worth noting that the choice of parametrization of total mass density slope $\gamma$ has a significant influence on cosmological constraints \cite{Qi:2018aio, Wang:2019yob, Chen:2018jcf}. In this paper, for the parametrization of $\gamma$, we take into account the dependencies of redshift and surface mass density, which introduces two additional lens model parameters, $\gamma_z$ and $\gamma_s$. To eliminate the uncertainties of constrained cosmological parameters by introducing two additional lens model parameters, it is required to know how both the masses and sizes of galaxies change with time, which is poorly understood. Besides, the influence of the prior of $\beta_{\rm{ani}}$ cannot be ignored \cite{Chen:2018jcf}, but its measurement is not very accurate at present. The lack of understanding of all these lens model parameters yields additional uncertainties in the estimation of cosmological parameters. As future massive surveys observe more and more SGL samples, a more accurate phenomenological model for lens galaxies could be characterized, which will greatly improve the constraint on cosmology using SGL data.

On the other hand, an abundant SGL sample with accurate measurements will be observed in the LSST era. We also make a forecast for the constraints on cosmological parameters from 8000 SGL systems with well-measured velocity dispersion and 55 SGL systems with well-measured time delay. We find that the TD observation with a small sample size can only constrain $H_0$ strictly, but has a very weak ability to constrain other cosmological parameters. The VD observation with a large sample size could be highly complementary with the TD observation to significantly break the parameter degeneracies. For example, for the $w$CDM model, the joint data analysis gives $\sigma(H_0)=0.57\Mpc$, $\sigma(\Omega_m)=0.005$ and $\sigma(w)=0.05$, which are comparable with the results from Planck 2018 TT,TE,EE+lowE+lensing+SNe+BAO, and meet the standard of precision cosmology. We conclude that the observations of SGL will become a useful late-universe probe for precisely measuring cosmological parameters and exploring the nature of dark energy.

\acknowledgments{This work was supported by the National Natural Science Foundation of China (Grants Nos. 11975072, 11835009, 11875102, and 11690021), the Liaoning Revitalization Talents Program (Grant No. XLYC1905011), the Fundamental Research Funds for the Central Universities (Grants Nos. N2105014 and N2005030), the National Program for Support of Top-Notch Young Professionals (Grant No. W02070050), the National 111 Project of China (Grant No. B16009), and the Science Research Grants from the China Manned Space Project (Grant No. CMS-CSST-2021-B01). }


\begin{adjustwidth}{-\extralength}{0cm}

\reftitle{References}


\bibliography{tdvd}


%


\end{adjustwidth}
\end{document}